\documentclass[oneside]{article}
\usepackage[T1]{fontenc}
\usepackage{authblk}
\usepackage{float}
\usepackage{amsmath}
\usepackage{graphicx}
\usepackage{xfrac}
\usepackage[english]{babel}
\usepackage{lmodern}
\usepackage[T1]{fontenc}
\usepackage{xcolor}
\usepackage[latin9]{inputenc}
\usepackage{tikz}
\usepackage{mathtools}
\usepackage{graphicx}
\usepackage{esint}
\usepackage[unicode=true,
 bookmarks=false,
breaklinks=false,pdfborder={0 0 1},backref=false,colorlinks=false]
 {hyperref}
\usepackage[a4paper, total={210mm,297mm},margin=2.0cm]{geometry}
\usepackage{pgfplots}
\usepackage{color}
\usepackage{xspace}
\usepackage{caption}
\usepackage{subcaption}
\usepackage[export]{adjustbox}
\usepackage{multirow}
\usepackage{lmodern}
\usepackage{siunitx}
\usepackage{booktabs}
\usetikzlibrary{arrows,decorations.markings}
\pgfplotsset{compat=newest}
\usepackage{pgfplots}
\newlength\figureheight 
\newlength\figurewidth 

\title{Fluid flow around NACA 0012 airfoil at low-Reynolds numbers with hybrid lattice Boltzmann method}

\author[1]{G. Di Ilio \thanks{Electronic address: \texttt{giovanni.diilio@unicusano.it}; Corresponding author}}
\author[1]{D. Chiappini}
\author[2]{S. Ubertini}
\author[3]{G. Bella}
\author[4]{S. Succi}

\affil[1]{University of Rome ``Niccol\`{o} Cusano'', Via don Carlo Gnocci 3, 00166, Rome, Italy}
\affil[2]{University of Tuscia, Largo dell'Universit\`{a} snc, 01100, Viterbo, Italy}
\affil[3]{University of Rome ``Tor Vergata'', Via del Politecnico 1, 00133, Rome, Italy}
\affil[4]{Istituto Applicazioni Calcolo, CNR, Via dei Taurini 19, 00185, Rome, Italy}

\date{\today}

\begin{document}

\maketitle

\begin{abstract}
We simulate the two-dimensional fluid flow around National Advisory Committee for Aeronautics (NACA) 0012 airfoil using a hybrid lattice Boltzmann method (HLBM), which combines the standard lattice Boltzmann method with an unstructured finite-volume formulation. The aim of the study is to assess the numerical performances and the robustness of the computational method. 
To this purpose, after providing a convergence study to estimate the overall accuracy of the method, we analyze the numerical solution for different values of the angle of attack at a Reynolds number equal to $10^3$. Subsequently, flow fields at Reynolds numbers up to $10^4$ are computed for a zero angle of attack configuration.
A grid refinement scheme is applied to the uniformly spaced component of the overlapping grid system to further enhance the numerical efficiency of the model.
The results demonstrate the capability of the HLBM to achieve high accuracy near solid curved walls, thus providing a viable alternative in the realm of off-lattice Boltzmann methods based on body-fitted mesh.
\end{abstract}

\section{Introduction}

The lattice Boltzmann methods (LBM) \cite{Succi,Chen,Succi2,Benzi,Higuera,McNamara} are a relatively recent approach to computational fluid dynamics (CFD), which has been proven to be successful in a broad range of applications, from turbulence \cite{Karlin,Bosch,Dorschner2017}, to multiphase and free-surface flows \cite{Falcucci2009,DiFrancesco}, as well as to non-Newtonian flows \cite{DiIlio2016}, fluid-structure interaction problems \cite{DeRosis}, porous media \cite{Chiappini,Chiappini2015,Zarghami2014a} and beyond.
The original formulation of LBM is based on uniform cartesian grids. Such a feature makes the algorithm particularly easy to implement and the numerical scheme extremely efficient; moreover, LBM algorithms are characterized by an high level of scalability on parallel processing systems. Despite these appealing peculiarities, the uniform space discretization affects the capability of LBM to solve problems involving complex geometries and multi-scale problems. This may constitute an obstacle for many typical practical engineering applications.
To overcome this issue and extend the applicability of LBM, many variants of the original LB scheme have been proposed in literature. First chronological appearances of finite-difference based schemes are those of Higuera and Succi \cite{Higuera2} and Nannelli and Succi \cite{Nannelli}, where LBM is applied on irregular lattices. Numerical implementations on unstructured grids have been first introduced by Peng et al. \cite{Peng,Peng2} and Xi et al. \cite{Xi} who proposed a cell-vertex finite-volume approach to numerically solve the integral form of the lattice Boltzmann equation. Other finite-volume techniques are those presented by Patil and Lakshmisha \cite{Patil2009,Patil2012,Patil2013}, Chen and Schaefer \cite{Leitao}, and some of the authors of the present paper \cite{Ubertini,Ubertini2,Ubertini4,ZarghamiBiscarini,Zarghami2013}.
A different strategy to solve the lattice Boltzmann equation on non-uniform and unstructured meshes is based on Galerkin-type finite element methods and it has been introduced in \cite{Lee1,Lee2,Min,Patel}. Other approaches are the general characteristics based off-lattice Boltzmann scheme proposed by Bardow et al. \cite{Bardow}, the discrete unified gas kinetic scheme (DUGKS) \cite{Zhu} and the semi-lagrangian approach of Kramer et al. \cite{Kramer}. In \cite{Rao} a complete comparative study is performed to evaluate the performance of several explicit off-lattice Boltzmann methods.
\\
Although the mentioned schemes enhance significantly the geometrical flexibility of LBM, in general, they increase the complexity of the method itself, partially losing some key advantages. Moreover, their computational efficiency is considerably lower than that of traditional LBM. To overcome this limitation, we recently proposed an hybrid lattice Boltzmann approach (HLBM) based on the coupling between a uniform grid model and an unstructured body-fitted grid model \cite{DiIlio2017}. The HLBM retains the outstanding advantages of traditional LBM while bringing the flexibility of an unstructured discretization scheme. 
\\
In this work, we present a numerical investigation on the unsteady behavior of the fluid flow over a stationary NACA 0012 airfoil in the low Reynolds numbers regime, with the aim of further analyze the capabilities of the HLBM. Results demonstrate that the new method is robust and efficient and, for this reason, it can be regarded as a viable approach to fluid dynamic problems involving curved geometries.

\section{The hybrid strategy}

In this section, we briefly recall the principles and the key points of the HLBM.
\\
The HLBM is constructed on the overlapping between a uniform cartesian grid, where the traditional LBM is applied, with an unstructured grid made of triangular elements, where a finite-volume lattice Boltzmann formulation is applied. Both the numerical methods are based on the single-time relaxation approach.
The structured uniform grid covers the whole computational domain, while the unstructured one is defined in a confined region, that is, around the solid body.
\\
For a distribution function $f_i(\textbf{x},t)$, representing the probability of finding a fluid particle at position $\textbf{x}$ and time \textit{t} that is moving along the \textit{i}-th lattice direction with a discrete speed $\textbf{c}_i$, the lattice Boltzmann equation reads as follows:

\begin{equation}
f_i\left(\textbf{x}+\textbf{c}_i\Delta t_\textbf{s},t+\Delta t_\textbf{s}\right)-f_i\left(\textbf{x},t\right) = -\dfrac{\Delta t_\textbf{s}}{\tau_\textbf{s}}\left[f_i\left(\textbf{x},t\right) - f_i^{eq}\left(\textbf{x},t\right)\right],
\label{eq:LBE}
\end{equation}

Equation \ref{eq:LBE} is solved numerically on a uniform Cartesian grid, with $\tau_\textbf{s}$ and $\Delta t_\textbf{s}$ being defined as the relaxation time and the time step, respectively, related to the 'structured' scheme.
The equilibrium distribution function $f_i^{eq}(\textbf{x},t)$ is calculated as follows:

\begin{equation}
f_i^{eq}\left(\textbf{x},t\right) = w_i\rho\left(\textbf{x},t\right) \bigg\{1+\dfrac{\textbf{c}_i\cdot\textbf{u}\left(\textbf{x},t\right)}{c_s^2} +\dfrac{\left[\textbf{c}_i\cdot\textbf{u}\left(\textbf{x},t\right)\right]^2}{2c_s^4}- \dfrac{\left[\textbf{u}\left(\textbf{x},t\right)\right]^2}{2c_s^2}\bigg\}
\label{eq:local_eq}
\end{equation}

where $c_s$ is the lattice speed of sound, the parameters $w_i$ are a set of weights normalized to unity, and $\rho\left(\textbf{x},t\right)$ and $\textbf{u}\left(\textbf{x},t\right)$ are the macroscopic variables, namely fluid density and velocity, respectively.
For an ideal incompressible fluid flow, equation \ref{eq:LBE} reproduces the Navier-Stokes equations, with pressure $p = \rho c_s^2$ and the kinematic viscosity being defined as $\nu = c_s^2 (\tau_\textbf{s}-\frac{\Delta t_\textbf{s}}{2})$.
\\
The finite-volume lattice Boltzmann method adopted on the unstructured grid is a cell-vertex type scheme. The following equation applies at each node \textit{P} of the grid:

\begin{equation}
f_i(P,t+\Delta t_\textbf{u}) = f_i(P,t)+\Delta t_\textbf{u}\sum_{k=0}^{K'}S_{i k}f_i(P_k,t)-\frac{\Delta t_\textbf{u}}{\tau_\textbf{u}}\sum_{k=0}^{K'}C_{i k}[f_i(P_k,t)-f_i^{eq}(P_k,t)].
\label{eq:ulbe}
\end{equation}

Details for derivation of such equation are provided in \cite{Ubertini,DiIlio2017}, therefore are not reported here.
In equation \ref{eq:ulbe}, $\tau_\textbf{u}$ and $\Delta t_\textbf{u}$ are the relaxation time and the time step, respectively, related to the 'unstructured' scheme, $k=0$ denotes the pivotal node $P$ and the summations run over the nodes $P_k$ connected to $P$; the quantities $S_{i k}$ and $C_{i k}$ represent the streaming and collisional matrices of the \textit{i}-th population related to the \textit{k}-th node, respectively; the equilibrium distribution function is defined by equation \eqref{eq:local_eq}. The theoretical kinematic viscosity associated with this scheme is $\nu = c_s^2 \tau_\textbf{u}$ \cite{Ubertini}.
\\
Within the hybrid framework, equations \eqref{eq:LBE} and \eqref{eq:ulbe} are solved consistently in time, with the exchange of information taking place at predefined interpolation nodes, which are placed over the overlapping grids region, in terms of both macroscopic variables and distribution functions.
The following equations describe the post-collision for the structured and unstructured interpolation nodes, respectively:

\begin{equation}
\tilde{f_i^{\textbf{s}}} = f_i^{eq,\textbf{u}} + \frac{2\left(\tau_\textbf{s}-\Delta t_\textbf{s}\right)}{2\tau_\textbf{s}-\Delta t_\textbf{s}} f_i^{neq,\textbf{u}}
\label{eq:postcollision2_s}
\end{equation}

\begin{equation}
\tilde{f_i^{\textbf{u}}} = f_i^{eq,\textbf{s}} + \left(1-\frac{\Delta t_\textbf{s}}{2\tau_\textbf{s}}\right) f_i^{neq,\textbf{s}}+\Delta t_\textbf{u} \sum_{k=0}^{K'} S_{i k} \left[f_{i k}^{eq,\star}+f_{i k}^{neq,\star}\right]-\frac{\Delta t_\textbf{u}}{\tau_\textbf{u}} \sum_{k=0}^{K'} C_{i k} f_{i k}^{neq,\star}
\label{eq:postcollision2_u}
\end{equation}

where superscripts \textbf{s} and \textbf{u} refer to the 'structured' and the 'unstructured' nodes, respectively, and $f_{i}^{neq}$ is the non-equilibrium distribution function.
The quantities $f_{i k}^{eq,\star}$ and $f_{i k}^{neq,\star}$ in the summation terms of equation \eqref{eq:postcollision2_u} are defined as follows:

\begin{equation}
f_{i k}^{eq,\star} = f_{i k}^{eq,\textbf{s}},
\qquad
f_{i k}^{neq,\star} = \left(1-\frac{\Delta t_\textbf{s}}{2\tau_\textbf{s}}\right)f_{i k}^{neq,\textbf{s}}
\end{equation}

or

\begin{equation}
f_{i k}^{eq,\star} = f_{i k}^{eq,\textbf{u}},
\qquad
f_{i k}^{neq,\star} = f_{i k}^{neq,\textbf{u}}
\end{equation}

depending on whether the \textit{k}-th node is an interpolation node or not, respectively. The distribution functions $f_i^{eq,\textbf{u}}$ and $f_i^{neq,\textbf{u}}$ of the right-hand side of equation \eqref{eq:postcollision2_s} and $f_i^{eq,\textbf{s}}$ and $f_i^{neq,\textbf{s}}$ of the right-hand side of equation \eqref{eq:postcollision2_u} are evaluated by interpolation procedure.

\section{The interpolation scheme}
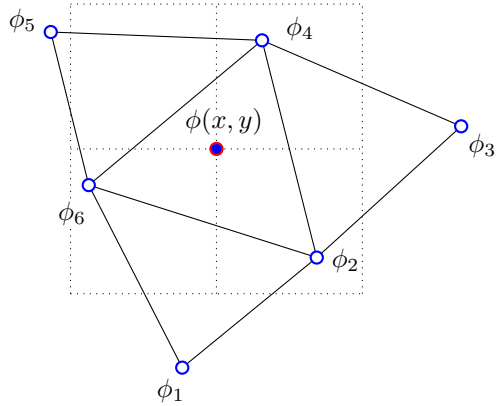
\begin{figure}[H]
	\centering
	\begin{tikzpicture}

\draw [dotted] (-0.72,-0.72) -- (3.12,-0.72) ;
\draw [dotted] (-0.72,1.2) -- (3.12,1.2) ;
\draw [dotted] (-0.72,3.12) -- (3.12,3.12) ;
\draw [dotted] (-0.72,-0.72) -- (-0.72,3.12) ;
\draw [dotted] (1.2,-0.72) -- (1.2,3.12) ;
\draw [dotted] (3.12,-0.72) -- (3.12,3.12) ;

\draw (1.8,2.64) -- (-0.48,0.72) ;
\draw (1.8,2.64) -- (2.52,-0.24) ;
\draw (2.52,-0.24) -- (-0.48,0.72) ;
\draw (2.52,-0.24) -- (4.42,1.5) ;
\draw (1.8,2.64) -- (4.42,1.5) ;
\draw (-0.98,2.75) -- (1.8,2.64) ;
\draw (-0.98,2.75) -- (-0.48,0.72) ;
\draw (0.75,-1.7) -- (-0.48,0.72) ;
\draw (0.75,-1.7) -- (2.52,-0.24) ;

\draw [red, thick, fill=blue] (1.2,1.2) circle [radius=0.08] ;

\draw [blue, thick, fill=white] (1.8,2.64) circle [radius=0.08] ;
\draw [blue, thick, fill=white] (-0.48,0.72) circle [radius=0.08] ;
\draw [blue, thick, fill=white] (2.52,-0.24) circle [radius=0.08] ;
\draw [blue, thick, fill=white] (4.42, 1.5) circle [radius=0.08] ;
\draw [blue, thick, fill=white] (-0.98, 2.75) circle [radius=0.08] ;
\draw [blue, thick, fill=white] (0.75,-1.7) circle [radius=0.08] ;

\draw [white, fill=white] (1.3,1.56) circle [radius=0.15] ;
\node at (1.3,1.56) {$\phi(x,y)$};

\node at (0.6,-2) {$\phi_1$};
\node at (4.7,1.25) {$\phi_3$};
\node at (-1.35,2.9) {$\phi_5$};
\draw [white, fill=white] (-0.7,0.35) circle [radius=0.15] ;
\node at (-0.7,0.35) {$\phi_6$};
\draw [white, fill=white] (2.9,-0.28) circle [radius=0.15] ;
\node at (2.9,-0.28) {$\phi_2$};
\draw [white, fill=white] (2.3,2.8) circle [radius=0.15] ;
\node at (2.3,2.8) {$\phi_4$};

\end{tikzpicture}
	\caption{Interpolation scheme representation for the structured nodes. The generic variable $\phi(x,y)$ is approximated by a linear combination of the shape functions defined on the six unstructured donor nodes. \label{fig:interpolation_st}}
\end{figure}

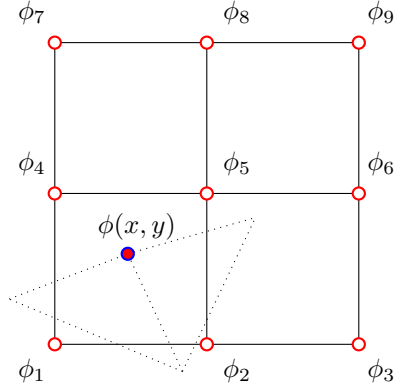
\begin{figure}[H]
	\centering
	\begin{tikzpicture}

\draw (0,0) -- (2,0) ;
\draw (0,2) -- (2,2) ;
\draw (0,0) -- (0,2) ;
\draw (2,0) -- (2,2) ;
\draw (2,0) -- (4,0) ;
\draw (4,0) -- (4,2) ;
\draw (4,4) -- (4,2) ;
\draw (4,4) -- (2,4) ;
\draw (2,4) -- (2,2) ;
\draw (4,2) -- (2,2) ;
\draw (0,4) -- (0,2) ;
\draw (0,4) -- (2,4) ;
\draw [red, thick, fill=white] (0,0) circle [radius=0.08] ;
\draw [red, thick, fill=white] (0,2) circle [radius=0.08] ;
\draw [red, thick, fill=white] (2,0) circle [radius=0.08] ;
\draw [red, thick, fill=white] (2,2) circle [radius=0.08] ;
\draw [red, thick, fill=white] (0,4) circle [radius=0.08] ;
\draw [red, thick, fill=white] (4,0) circle [radius=0.08] ;
\draw [red, thick, fill=white] (4,2) circle [radius=0.08] ;
\draw [red, thick, fill=white] (2,4) circle [radius=0.08] ;
\draw [red, thick, fill=white] (4,4) circle [radius=0.08] ;

\draw [dotted] (0.96,1.2) -- (2.64,1.68) ;
\draw [dotted] (-0.6,0.6) -- (0.96,1.2) ;
\draw [dotted] (1.68,-0.36) -- (2.64,1.68) ;
\draw [dotted] (1.68,-0.36) -- (0.96,1.2) ;
\draw [dotted] (1.68,-0.36) -- (-0.6,0.6) ;

\draw [blue, thick, fill=red] (0.96,1.2) circle [radius=0.08] ;

\node at (1.08,1.56) {$\phi(x,y)$};

\node at (-0.3,-0.35) {$\phi_1$};
\node at ( 2.4,-0.35) {$\phi_2$};
\node at ( 4.3,-0.35) {$\phi_3$};
\node at (-0.3, 2.4 ) {$\phi_4$};
\node at ( 2.4, 2.4 ) {$\phi_5$};
\node at ( 4.3, 2.4 ) {$\phi_6$};
\node at (-0.3, 4.4 ) {$\phi_7$};
\node at ( 2.4, 4.4 ) {$\phi_8$};
\node at ( 4.3, 4.4 ) {$\phi_9$};

\end{tikzpicture}
	\caption{Interpolation scheme representation for the unstructured nodes. The generic variable $\phi(x,y)$ is approximated by a linear combination of the shape functions defined on the nine structured donor nodes. \label{fig:interpolation_un}}
\end{figure}

For traditional multi-domain grid refinement, the information communication on grid transitions significantly affects the accuracy of the numerical scheme \cite{Lagrava}. Similarly, in the HLBM, the exchange of information between structured and unstructured grid is of crucial importance from the accuracy point of view.
Here we describe the interpolation scheme used for reconstructing both the macroscopic variables and the distribution functions from one grid to the other. Later, we will provide a convergence study, in order to assess the overall accuracy of the present HLBM.
\\
The hybrid grid presents interpolation nodes of two kinds: structured and unstructured. The quantities on the structured interpolation nodes are reconstructed by considering four triangular elements of the unstructured mesh, as depicted in Figure \ref{fig:interpolation_st}.
A set of 2nd order shape functions is defined on the six unstructured nodes surrounding the structured interpolation node. Therefore, the polygon composed of the four selected triangles represents a quadratic element, and the generic variable $\phi(x,y)$ on the structured node is approximated by a linear combination of the shape functions $\psi_k$, as follows:
\begin{equation}
\phi(x,y)=\sum\limits_{k=1}^{6}\phi_k\psi_k(x,y).
\label{eq:interpolation_st}
\end{equation}
Similarly, in order to reconstruct the quantities on the unstructured interpolation nodes, we select a square region, which is composed of four structured elements, as depicted in Figure \ref{fig:interpolation_un}.
On the nine structured nodes surrounding the unstructured interpolation node we define a set of 2nd order shape functions. Therefore, the generic variable $\phi(x,y)$ on the unstructured node is evaluated by means of a linear combination of these shape functions, as follows:
\begin{equation}
\phi(x,y)=\sum\limits_{k=1}^{9}\phi_k\psi_k(x,y).
\label{eq:interpolation_un}
\end{equation}

\section{Problem setup}
\begin{figure}[H]
	\centering
	\includegraphics[width = 0.8\columnwidth]{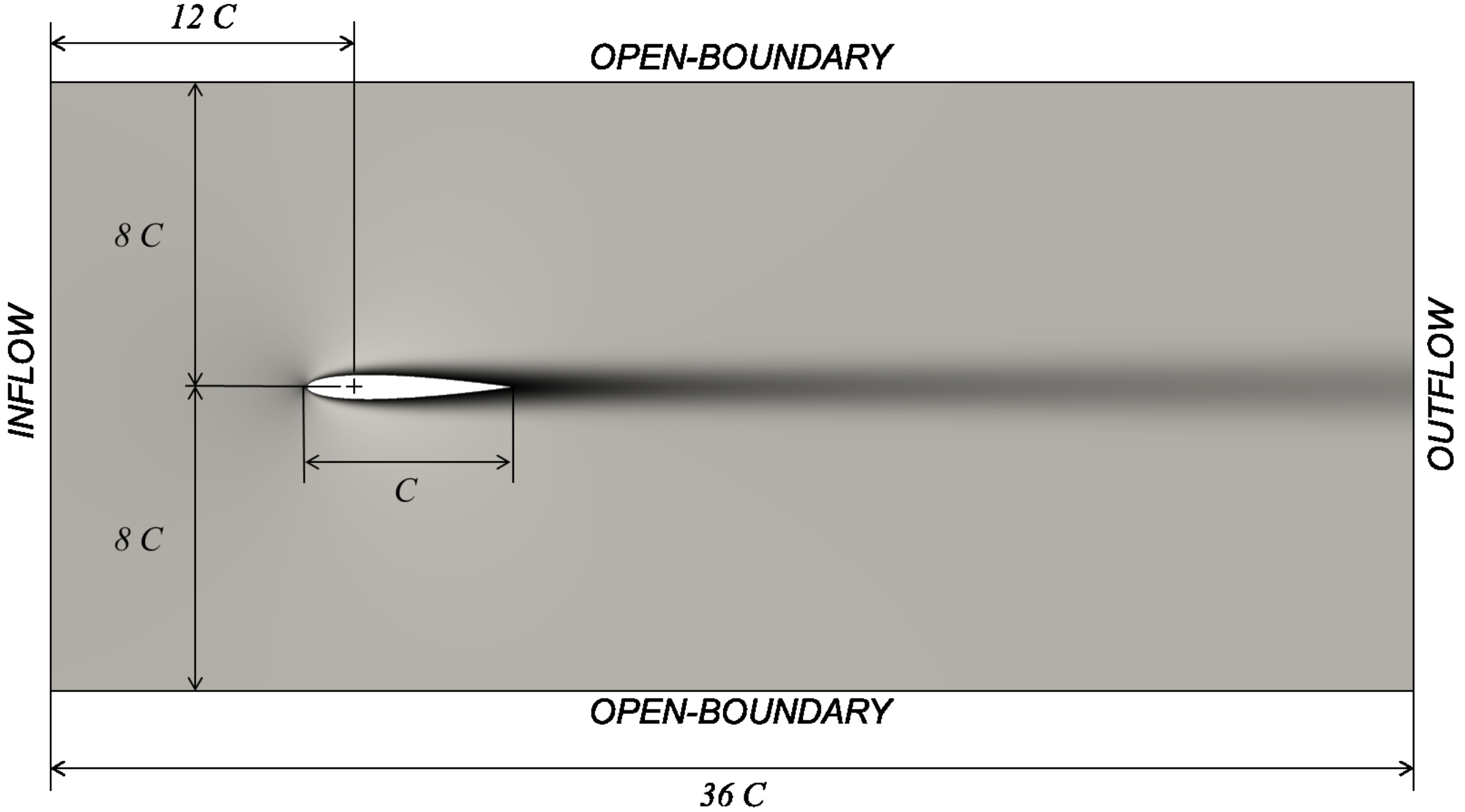}
	\caption{Domain configuration for simulation of fluid flow around NACA 0012 airfoil. 
	\label{fig:domain_NACA}}
\end{figure}

\begin{figure}[H]
	\centering
	\includegraphics[width = 0.8\columnwidth]{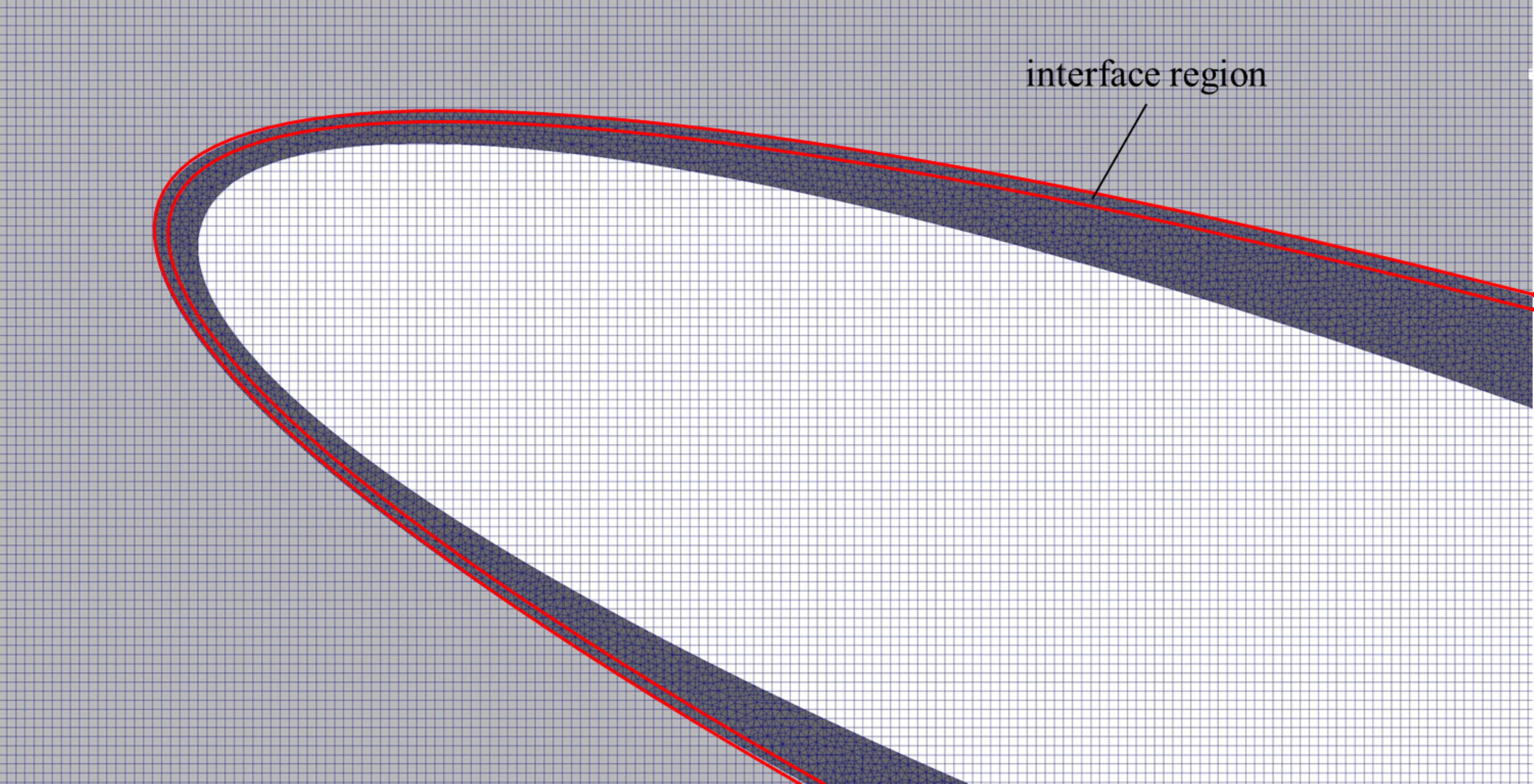}
	\caption{Detail of the hybrid mesh around NACA 0012 airfoil. The interface region is defined as the region between the two red curves.
	\label{fig:NACA_interface}}
\end{figure}

In this study, we simulate a laminar flow past a NACA 0012 airfoil, which is characterized by a symmetric profile whit 12$\%$ thickness to chord length ratio.
\\
NACA airfoils have been extensively used for the validation of numerical schemes, thanks to the availability of experimental data for several different profiles.
Here, with the aim of analyzing the influence of the angle of attack on the development and the evolution of vortices from the airfoil surface, a parametric study is conducted at Re = 1000 for a wide range of angles of attack. Moreover, flow fields are computed at higher values of Reynolds number, specifically Re = 2000, 5000 and 10000, at zero angle of attack.
\\
The geometry of the two-dimensional computational domain adopted for the present analysis is illustrated in Figure \ref{fig:domain_NACA} where $C$ indicates the chord length. The airfoil is placed symmetrically with respect to bottom and top boundaries.
In order to further enhance the computational efficiency of the hybrid method, we implement the grid refinement technique of Filippova and Hanel \cite{Filippova} within the structured sub-model scheme.
In particular, we consider four structured refinement levels. Each structured level is constituted by a rectangular box, which represents a resolution region in which the structured mesh has uniform discretization.
The grid refinement is performed by dividing the lattice spacing by a factor of two, with the inner boxes having higher resolution over the outer box in which they are contained. The innermost structured region, which is characterized by a unitary lattice spacing, overlaps the unstructured mesh surrounding the airfoil.
\\
No-slip wall boundary condition is applied at the airfoil surface, uniform velocity and zero pressure gradient are set at the inlet, fixed pressure value and zero velocity gradient at the outlet, and open-boundary conditions are applied to the top and bottom walls of the domain.
\\
The unstructured mesh surrounding the solid body is constituted by an ellipse. This choice is purely arbitrary, and it has been taken after testing other elementary shapes and assessing the non-dependency of the results from the unstructured mesh configuration.
Figure \ref{fig:NACA_interface} shows a detail for a generic configuration of the selected unstructured discretization around the airfoil. In the figure, the overlapping area between the finest structured grid and the unstructured grid is represented as the region enclosed by two red lines. The interpolation nodes, which are responsible of the exchange of information between the two grid structures, are placed along this region. Also the internal fictitious curve defining such an interface region is constituted by an ellipse.
\\

\section{Convergence study}
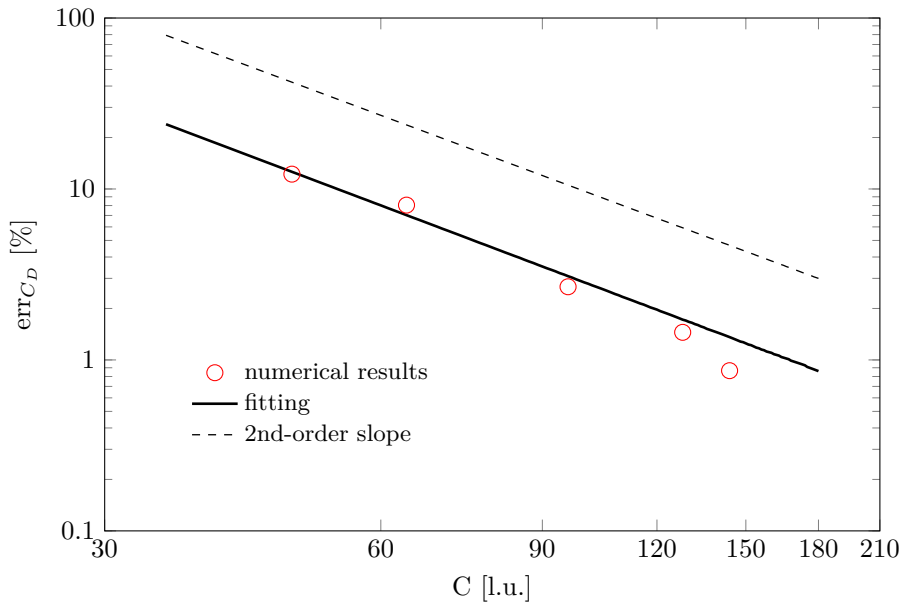
\begin{figure}[H]
	\centering 
	\setlength\figureheight{0.4\columnwidth}
	\setlength\figurewidth{0.6\columnwidth}
	\begin{tikzpicture}

\begin{loglogaxis}[%
width=\figurewidth,
height=\figureheight,
scale only axis,
log ticks with fixed point,
xmin=30,
xmax=210,
xlabel={C [l.u.]},
xtick={30, 60, 90, 120, 150, 180, 210},
ymin=0.1,
ymax=100,
ylabel={err$_{C_D}$ [$\%$]},
legend style={draw=none, at={(0.1,0.35)},anchor=north west,legend cell align=left,font = \small}
]

\addplot [color=red, only marks, mark size=3.0pt, mark=o]
  table[row sep=crcr]{%
48	12.2116 \\
64	 8.0465 \\
96	 2.6790 \\
128	 1.4503 \\
144	 0.8653 \\
};

\addplot [color=black,line width=1pt]
  table[row sep=crcr]{%
35	23.91	\\
36	22.59	\\
37	21.36	\\
38	20.24	\\
39	19.20	\\
40	18.24	\\
41	17.35	\\
42	16.52	\\
43	15.75	\\
44	15.03	\\
45	14.36	\\
46	13.74	\\
47	13.15	\\
48	12.60	\\
49	12.09	\\
50	11.60	\\
51	11.14	\\
52	10.71	\\
53	10.31	\\
54	9.92	\\
55	9.56	\\
56	9.22	\\
57	8.89	\\
58	8.59	\\
59	8.29	\\
60	8.02	\\
61	7.75	\\
62	7.50	\\
63	7.26	\\
64	7.03	\\
65	6.81	\\
66	6.61	\\
67	6.41	\\
68	6.22	\\
69	6.04	\\
70	5.86	\\
71	5.70	\\
72	5.54	\\
73	5.39	\\
74	5.24	\\
75	5.10	\\
76	4.96	\\
77	4.83	\\
78	4.71	\\
79	4.59	\\
80	4.47	\\
81	4.36	\\
82	4.25	\\
83	4.15	\\
84	4.05	\\
85	3.96	\\
86	3.86	\\
87	3.77	\\
88	3.69	\\
89	3.60	\\
90	3.52	\\
91	3.44	\\
92	3.37	\\
93	3.30	\\
94	3.22	\\
95	3.16	\\
96	3.09	\\
97	3.03	\\
98	2.96	\\
99	2.90	\\
100	2.84	\\
101	2.79	\\
102	2.73	\\
103	2.68	\\
104	2.63	\\
105	2.58	\\
106	2.53	\\
107	2.48	\\
108	2.43	\\
109	2.39	\\
110	2.34	\\
111	2.30	\\
112	2.26	\\
113	2.22	\\
114	2.18	\\
115	2.14	\\
116	2.11	\\
117	2.07	\\
118	2.03	\\
119	2.00	\\
120	1.97	\\
121	1.93	\\
122	1.90	\\
123	1.87	\\
124	1.84	\\
125	1.81	\\
126	1.78	\\
127	1.75	\\
128	1.72	\\
129	1.70	\\
130	1.67	\\
131	1.65	\\
132	1.62	\\
133	1.60	\\
134	1.57	\\
135	1.55	\\
136	1.52	\\
137	1.50	\\
138	1.48	\\
139	1.46	\\
140	1.44	\\
141	1.42	\\
142	1.40	\\
143	1.38	\\
144	1.36	\\
145	1.34	\\
146	1.32	\\
147	1.30	\\
148	1.28	\\
149	1.27	\\
150	1.25	\\
151	1.23	\\
152	1.22	\\
153	1.20	\\
154	1.18	\\
155	1.17	\\
156	1.15	\\
157	1.14	\\
158	1.12	\\
159	1.11	\\
160	1.10	\\
161	1.08	\\
162	1.07	\\
163	1.06	\\
164	1.04	\\
165	1.03	\\
166	1.02	\\
167	1.01	\\
168	0.99	\\
169	0.98	\\
170	0.97	\\
171	0.96	\\
172	0.95	\\
173	0.94	\\
174	0.93	\\
175	0.91	\\
176	0.90	\\
177	0.89	\\
178	0.88	\\
179	0.87	\\
180	0.86	\\
};

\addplot [color=black, dashed, line width=0.5pt]
  table[row sep=crcr]{%
35	79.25	\\
36	74.91	\\
37	70.91	\\
38	67.23	\\
39	63.83	\\
40	60.68	\\
41	57.75	\\
42	55.03	\\
43	52.50	\\
44	50.14	\\
45	47.94	\\
46	45.88	\\
47	43.95	\\
48	42.14	\\
49	40.43	\\
50	38.83	\\
51	37.32	\\
52	35.90	\\
53	34.56	\\
54	33.29	\\
55	32.09	\\
56	30.96	\\
57	29.88	\\
58	28.86	\\
59	27.89	\\
60	26.97	\\
61	26.09	\\
62	25.25	\\
63	24.46	\\
64	23.70	\\
65	22.98	\\
66	22.29	\\
67	21.63	\\
68	20.99	\\
69	20.39	\\
70	19.81	\\
71	19.26	\\
72	18.73	\\
73	18.22	\\
74	17.73	\\
75	17.26	\\
76	16.81	\\
77	16.37	\\
78	15.96	\\
79	15.56	\\
80	15.17	\\
81	14.80	\\
82	14.44	\\
83	14.09	\\
84	13.76	\\
85	13.44	\\
86	13.13	\\
87	12.83	\\
88	12.54	\\
89	12.26	\\
90	11.99	\\
91	11.72	\\
92	11.47	\\
93	11.22	\\
94	10.99	\\
95	10.76	\\
96	10.53	\\
97	10.32	\\
98	10.11	\\
99	9.91	\\
100	9.71	\\
101	9.52	\\
102	9.33	\\
103	9.15	\\
104	8.98	\\
105	8.81	\\
106	8.64	\\
107	8.48	\\
108	8.32	\\
109	8.17	\\
110	8.02	\\
111	7.88	\\
112	7.74	\\
113	7.60	\\
114	7.47	\\
115	7.34	\\
116	7.21	\\
117	7.09	\\
118	6.97	\\
119	6.86	\\
120	6.74	\\
121	6.63	\\
122	6.52	\\
123	6.42	\\
124	6.31	\\
125	6.21	\\
126	6.11	\\
127	6.02	\\
128	5.93	\\
129	5.83	\\
130	5.74	\\
131	5.66	\\
132	5.57	\\
133	5.49	\\
134	5.41	\\
135	5.33	\\
136	5.25	\\
137	5.17	\\
138	5.10	\\
139	5.02	\\
140	4.95	\\
141	4.88	\\
142	4.81	\\
143	4.75	\\
144	4.68	\\
145	4.62	\\
146	4.55	\\
147	4.49	\\
148	4.43	\\
149	4.37	\\
150	4.31	\\
151	4.26	\\
152	4.20	\\
153	4.15	\\
154	4.09	\\
155	4.04	\\
156	3.99	\\
157	3.94	\\
158	3.89	\\
159	3.84	\\
160	3.79	\\
161	3.75	\\
162	3.70	\\
163	3.65	\\
164	3.61	\\
165	3.57	\\
166	3.52	\\
167	3.48	\\
168	3.44	\\
169	3.40	\\
170	3.36	\\
171	3.32	\\
172	3.28	\\
173	3.24	\\
174	3.21	\\
175	3.17	\\
176	3.13	\\
177	3.10	\\
178	3.06	\\
179	3.03	\\
180	3.00	\\
};

\legend{numerical results, fitting, 2nd-order slope};

\end{loglogaxis}
\end{tikzpicture}%
	\caption{
	Convergence of the HLBM. The numerical results are represented by open circles; the solid thick line is the fitting curve. The second-order slope is also provided as a guide (dashed line).
	\label{fig:convergence}}
\end{figure}

In order to estimate the overall accuracy of the numerical method and to verify that the adopted interpolation scheme is compatible with the order of accuracy of the standard LBM we preliminary perform a convergence study.
\\
For this analysis we consider five hybrid meshes of different resolution. In particular, the following chord lengths are chosen: $C$ = 48, 64, 96, 128, 144. The average element-size at wall is the same for all the unstructured meshes. Considering the Reynolds number based on the free-stream velocity and the airfoil chord, the analysis is carried out at Re = 500. The angle of attack of the airfoil is set to $\alpha=0^{\circ}$. All the simulations are performed with same time-step for the unstructured routine.
\\
The results are provided in terms of percentage error for the drag coefficient, with reference to the value computed by \textit{XFOIL - Subsonic Airfoil Development System}. \textit{XFOIL} is an interactive program for the design and analysis of subsonic isolated airfoils, developed by Mark Drela at Massachusetts Institute of Technology \cite{Drela}.
In Figure \ref{fig:convergence} we show the deviation of the computed mean drag coefficient from the reference value.
We notice that the order of convergence of the present scheme is about equal to 2. This result indicates that the interpolation scheme preserves the second order accuracy of the standard LBM.

\section{Results and discussion}

\subsection{Effect of the angle of attack}
\label{AoAstudy}
\begin{figure}[H]
\centering
{\includegraphics[width = 0.35\columnwidth]{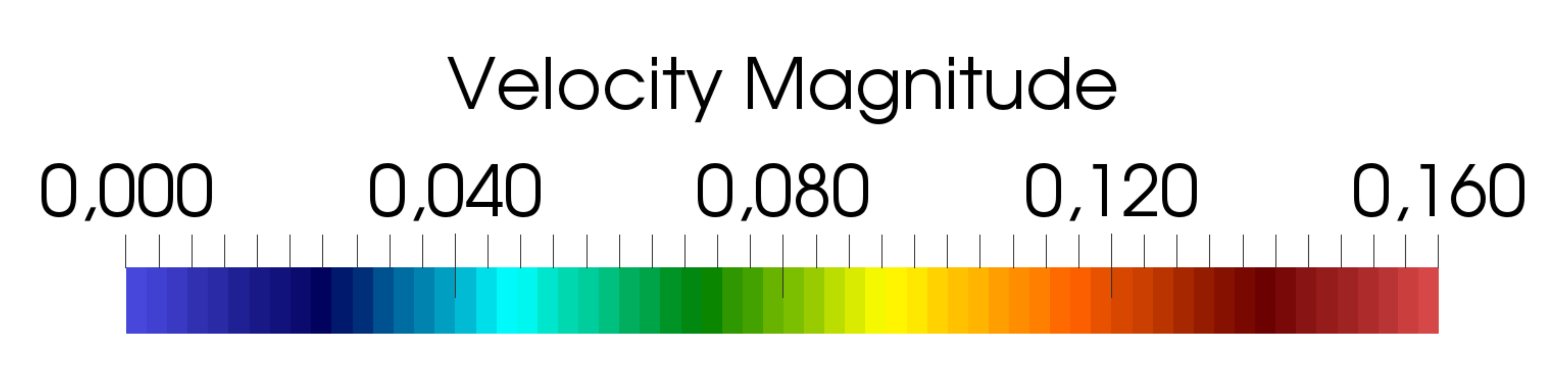}}
{\includegraphics[width = 0.35\columnwidth]{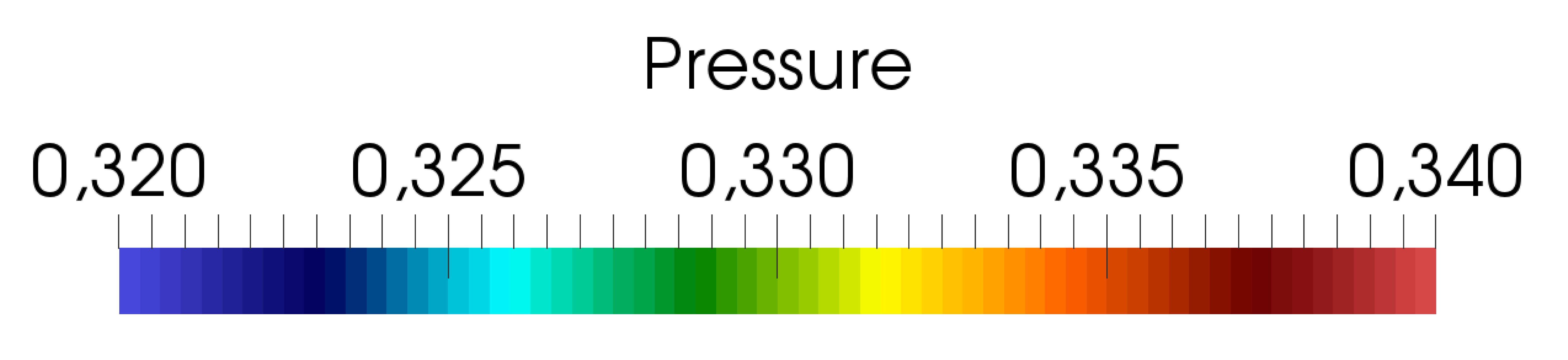}}
\subcaptionbox{}
{\includegraphics[width = 0.35\columnwidth]{./alfa00_vel.pdf}}
\subcaptionbox{}
{\includegraphics[width = 0.35\columnwidth]{./alfa00_pre.pdf}}
\subcaptionbox{}
{\includegraphics[width = 0.35\columnwidth]{./alfa08_vel.pdf}}
\subcaptionbox{}
{\includegraphics[width = 0.35\columnwidth]{./alfa08_pre.pdf}}
\subcaptionbox{}
{\includegraphics[width = 0.35\columnwidth]{./alfa10_vel.pdf}}
\subcaptionbox{}
{\includegraphics[width = 0.35\columnwidth]{./alfa10_pre.pdf}}
\subcaptionbox{}
{\includegraphics[width = 0.35\columnwidth]{./alfa20_vel.pdf}}
\subcaptionbox{}
{\includegraphics[width = 0.35\columnwidth]{./alfa20_pre.pdf}}
\subcaptionbox{}
{\includegraphics[width = 0.35\columnwidth]{./alfa28_vel.pdf}}
\subcaptionbox{}
{\includegraphics[width = 0.35\columnwidth]{./alfa28_pre.pdf}}
\caption{ Instantaneous velocity and pressure fields at Re = 1000: (a),(b): $\alpha=0^{\circ}$; (c),(d): $\alpha=8^{\circ}$; (e),(f): $\alpha=10^{\circ}$; (g),(h): $\alpha=20^{\circ}$; (i),(j): $\alpha=28^{\circ}$.
\label{fig:veldensField_NACA}}
\end{figure}

Simulations are performed for a range of angles of attack, $\alpha$, between $0^{\circ}$ and $29^{\circ}$. All the cases are studied at the fixed value of Re = 1000. For this analysis, the chord length of the airfoil is set to be equal to 512 lattice units. For each hybrid mesh, the total number of structured nodes is approximately equal to $3.3\cdot10^6$, while the number of unstructured nodes is significantly lower, that is about $2.2\cdot10^4$.
\\
Figure \ref{fig:veldensField_NACA} illustrates the instantaneous velocity and pressure fields for Re = 1000 at several angle of attack, namely: $\alpha$= $0^{\circ}$, $8^{\circ}$, $10^{\circ}$, $20^{\circ}$ and $28^{\circ}$.
For angles of attack lower than $4^{\circ}$, no flow separation is observed. However, beyond this particular value of $\alpha$, a recirculation region arise at the rear of the airfoil.
The unsteady vortex shedding phenomenon is observed at about $\alpha$ = $8^{\circ}$, as for $\alpha < 8^{\circ}$ the numerical solution converges to steady state. This evidence is in agreement with results available in literature \cite{Kurtulus,Huang,Ohmi,Liu}. 
A further increasing of the angle of attack gives rise to leading-edge vortexes and flow instabilities at the rear of the airfoil become larger. Then, starting from an angle of attack of 24 degrees, the vortex shedding becomes non-periodic.

\begin{figure}[H]
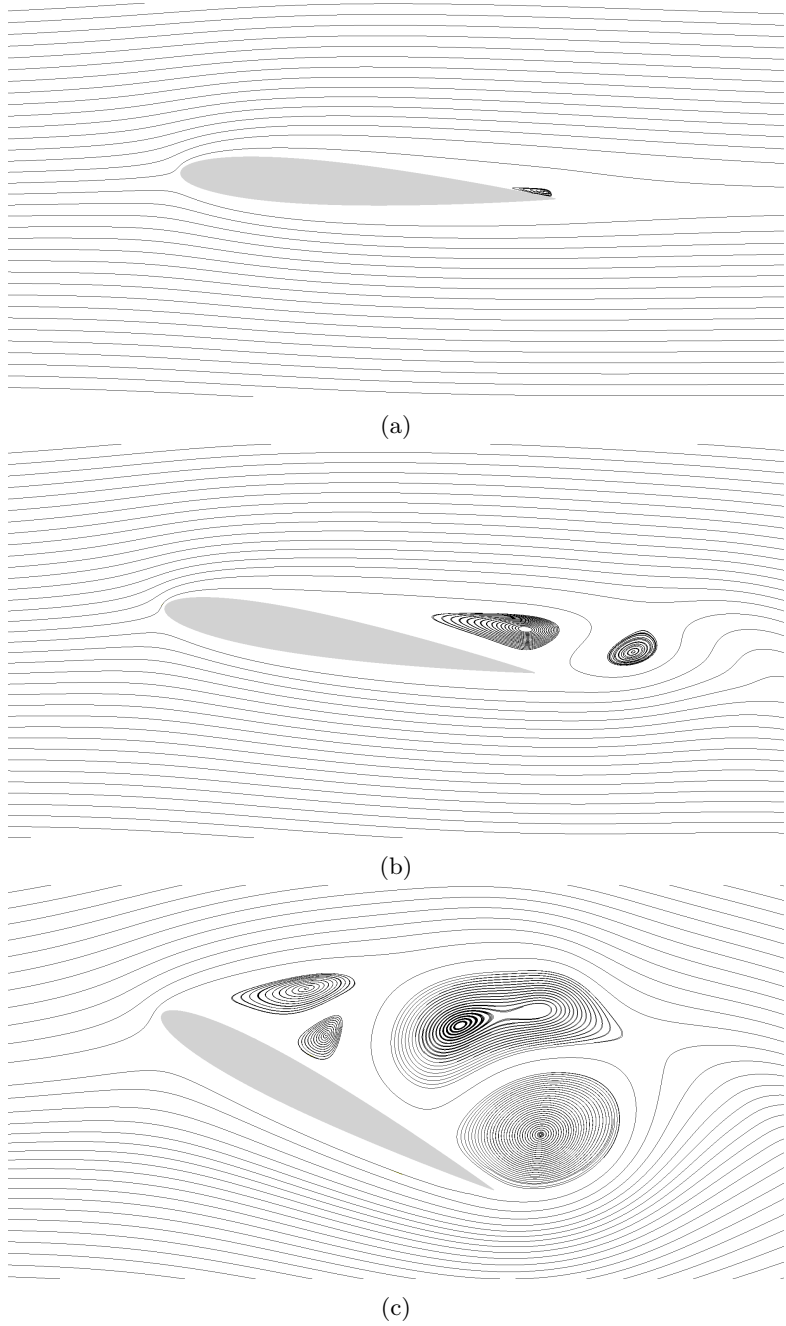

\centering
\subcaptionbox{}
{\includegraphics[width = 0.6\columnwidth]{./stream04.pdf}}
\subcaptionbox{}
{\includegraphics[width = 0.6\columnwidth]{./stream10.pdf}}
\subcaptionbox{}
{\includegraphics[width = 0.6\columnwidth]{./stream28.pdf}}
\caption{Instantaneous streamline patterns of the flow around NACA0012 airfoil at Re = 1000: (a): $\alpha=4^{\circ}$; (b): $\alpha=10^{\circ}$; (c): $\alpha=28^{\circ}$.
\label{fig:streamlines}}
\end{figure}

Figure \ref{fig:streamlines} shows the instantaneous streamline patterns for the fluid flow around the airfoil, at 4, 10 and 28 degrees, respectively.

\begin{figure}[H]
	\centering 
	\setlength\figureheight{0.4\columnwidth}
	\setlength\figurewidth{0.6\columnwidth}
	\input{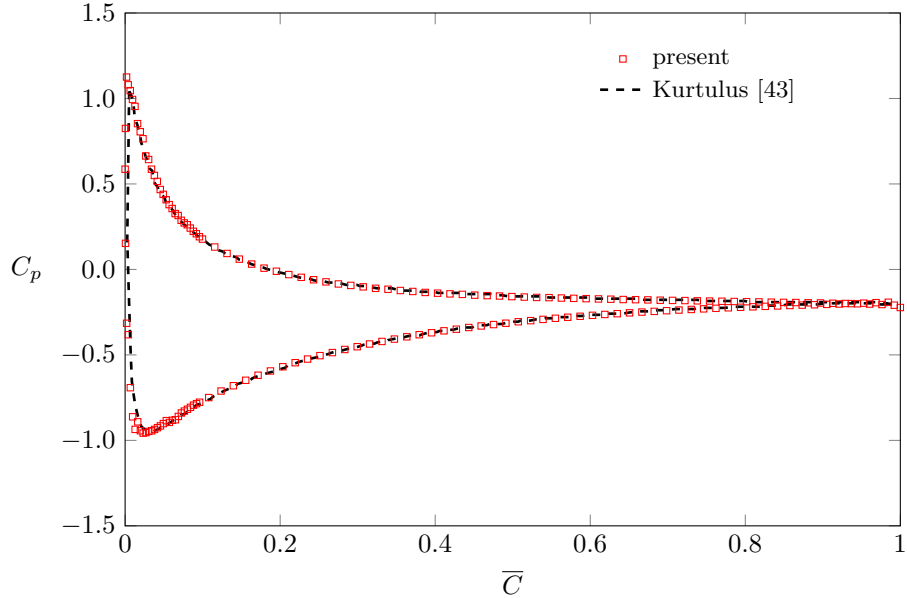}
	\caption{Mean pressure coefficient distribution over the upper and lower surfaces of the NACA 0012 airfoil at $\alpha=8^{\circ}$: comparison between present results and those reported in \cite{Kurtulus}.
	\label{fig:Cp_alfa8}}
\end{figure}

\begin{figure}[H]
	\centering 
	\setlength\figureheight{0.4\columnwidth}
	\setlength\figurewidth{0.6\columnwidth}
	\input{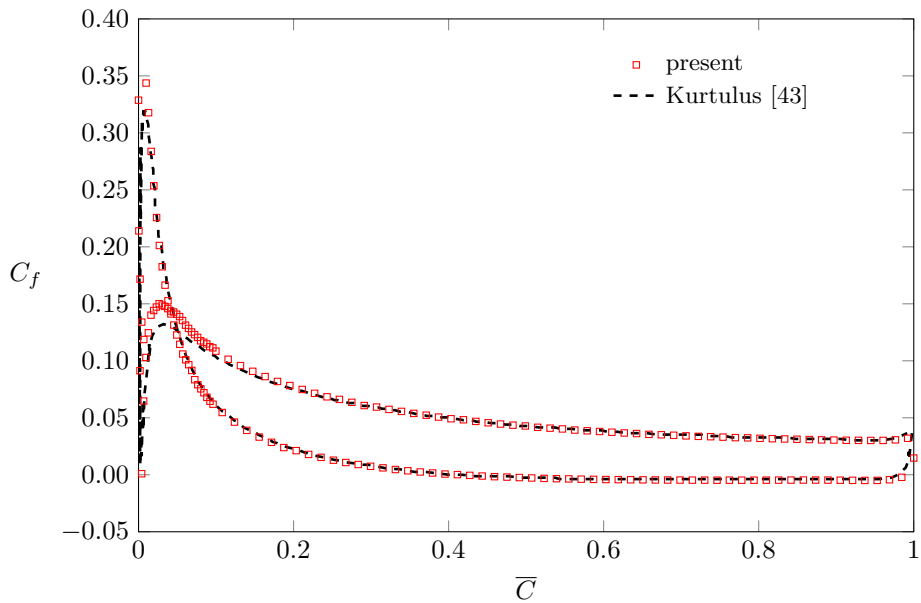}
	\caption{Mean skin friction coefficient distribution over the upper and lower surfaces of the NACA 0012 airfoil at $\alpha=8^{\circ}$: comparison between present results and those reported in \cite{Kurtulus}.
	\label{fig:Cf_alfa8}}
\end{figure}

\begin{figure}[H]
	\centering 
	\setlength\figureheight{0.4\columnwidth}
	\setlength\figurewidth{0.6\columnwidth}
	\begin{tikzpicture}

\begin{axis}[%
width=\figurewidth,
height=\figureheight,
scale only axis,
xmin=0,
xmax=30,
xtick={0, 5, 10, 15, 20, 25, 30},
xlabel={$\alpha [^{\circ}]$},
xmajorgrids,
ymin=0,
ymax=1,
ytick={0, 0.2, 0.4, 0.6, 0.8, 1},
ylabel={$C_D$},
ylabel style={rotate=-90},
ymajorgrids,
legend style={draw=none, at={(0.15,0.97)},anchor=north west,legend cell align=left,font = \small}
]

\addplot [color=black, line width=0.1pt, mark size=2.0pt, mark=*]
  table[row sep=crcr]{
0	0.119	\\
1	0.120	\\
2	0.121	\\
3	0.122	\\
4	0.125	\\
5	0.128	\\
6	0.132	\\
7	0.136	\\
8	0.141	\\
10	0.165	\\
12	0.196	\\
14	0.241	\\
16	0.307	\\
18	0.382	\\
20	0.442	\\
22	0.495	\\
24	0.571	\\
25	0.635	\\
26	0.678	\\
27	0.667	\\
28	0.680	\\
29	0.657	\\
};

\addplot [color=red, line width=0.1pt, mark size=2.0pt, mark=triangle]
  table[row sep=crcr]{
0.005	0.116	\\
1.000	0.116	\\
2.000	0.123	\\
3.000	0.123	\\
4.000	0.123	\\
5.000	0.123	\\
6.000	0.123	\\
7.000	0.131	\\
8.000	0.139	\\
9.000	0.150	\\
10.000	0.161	\\
10.971	0.174	\\
12.000	0.194	\\
13.000	0.222	\\
14.000	0.249	\\
15.066	0.275	\\
16.049	0.319	\\
17.000	0.352	\\
18.000	0.384	\\
19.162	0.413	\\
20.000	0.439	\\
21.000	0.453	\\
22.000	0.471	\\
23.095	0.536	\\
23.915	0.594	\\
24.899	0.652	\\
25.883	0.696	\\
26.864	0.681	\\
27.846	0.688	\\
29.000	0.739	\\
};

\legend{present study, Kurtulus \cite{Kurtulus}};

\end{axis}
\end{tikzpicture}
	\caption{Drag coefficient as a function of the angle of attack, at Re = 1000. Present results are compared with the numerical findings of Kurtulus \cite{Kurtulus}. \label{fig:NACA_CD}}
\end{figure}
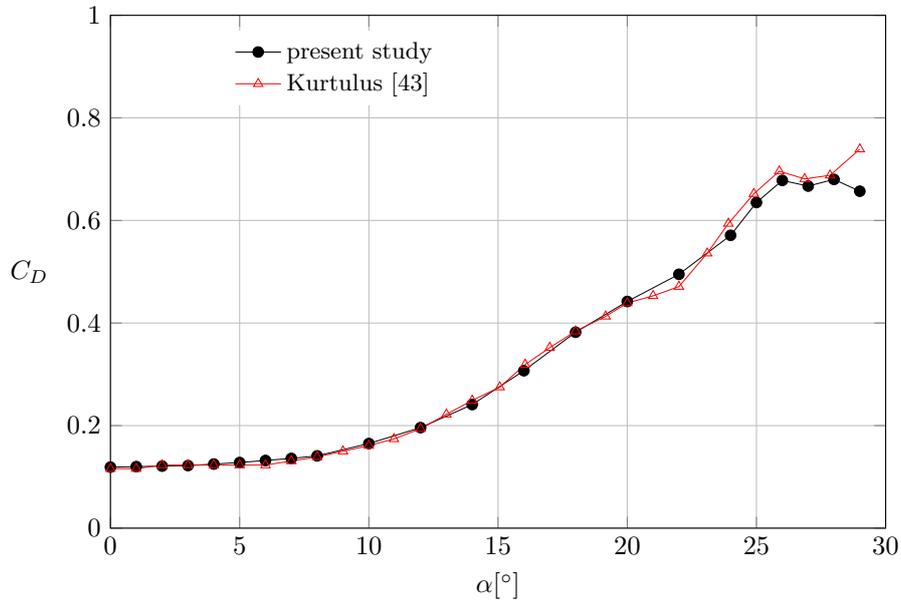

We note that, for $\alpha=4^{\circ}$, flow separation occurs and a recirculation area is observable at the rear of the airfoil. From Figure \ref{fig:streamlines}, it is also evident the shedding of the trailing-edge vortexes for $\alpha=10^{\circ}$.
\\
As the angle of attack is increased a large-scale flow separation on the surface of the airfoil takes place and the flow becomes highly unsteady. This behavior is already visible at $\alpha=28^{\circ}$.
\\
To asses the aerodynamic performance of the airfoil, drag and lift coefficients are calculated as follows:
\begin{equation}
C_D = \frac{F_D}{{\frac{1}{2}}{\rho_0}u_0^2 C},
\label{eq:dragNACA}
\end{equation}

\begin{equation}
C_L = \frac{F_L}{{\frac{1}{2}}{\rho_0}u_0^2 C},
\label{eq:liftNACA}
\end{equation}

where the quantities with subscript $0$ refer to the far field conditions, while $F_D$ and $F_L$ are the drag and lift force, respectively, namely the horizontal and the vertical components of the aerodynamic force acting on the solid body.
Moreover, we quantify the main features of the flow by computing the mean pressure coefficient and the skin friction coefficient as follows:

\begin{equation}
C_p = \frac{p-p_0}{{\frac{1}{2}}{\rho_0}u_0^2},
\label{eq:Cp}
\end{equation}

\begin{equation}
C_f = \frac{\tau_w}{{\frac{1}{2}}{\rho_0}u_0^2},
\label{eq:Cf}
\end{equation}

where $\tau_w$ denotes the mean wall shear stress.
We emphasize that the HLBM allows a direct computation of the forces on the surface of the body. In fact, thanks to the presence of the unstructured body-fitted mesh, both pressure and stress tensor are locally available at the computational nodes defining the geometry of the body. This implies that, to compute the aerodynamics variables, interpolation routines are not needed to be implemented.
\\
In Figures \ref{fig:Cp_alfa8} and \ref{fig:Cf_alfa8} we show the mean pressure coefficient and the mean skin friction coefficient, respectively, at $\alpha=8^{\circ}$, along the non-dimensional chord $\overline{C}$. The results are compared with those available from literature.
Despite a slight overestimation of the maximum value of the mean skin friction coefficient on both lower and upper surfaces of the airfoil, the overall agreement between our computed trends and those from literature is satisfactory. 
Also, the skin friction distribution provides a meaningful way to determine the separation point. In fact, this can be inferred from the location of zero skin friction with a negative gradient. The obtained result for $C_f$ indicates that the location of the separation point on the upper surface of the airfoil is approximately equal to 0.41$\overline{C}$, which is in line with the value provided by Kurtulus \cite{Kurtulus} (0.4$\overline{C}$).
\\
In Figures \ref{fig:NACA_CL} and \ref{fig:NACA_CD} the time-averaged drag and lift coefficients are reported as a function of the angle of attack, respectively. The figures shows also a comparison with literature data.

\begin{figure}[H]
	\centering 
	\setlength\figureheight{0.4\columnwidth}
	\setlength\figurewidth{0.6\columnwidth}
	\begin{tikzpicture}

\begin{axis}[%
width=\figurewidth,
height=\figureheight,
scale only axis,
xmin=0,
xmax=30,
xtick={0, 5, 10, 15, 20, 25, 30},
xlabel={$\alpha [^{\circ}]$},
xmajorgrids,
ymin=0,
ymax=1.4,
ytick={0, 0.2, 0.4, 0.6, 0.8, 1.0, 1.2, 1.4},
ylabel={$C_L$},
ylabel style={rotate=-90},
ymajorgrids,
legend style={draw=none, at={(0.08,0.98)},anchor=north west,legend cell align=left,font = \small}
]

\addplot [color=black, line width=0.1pt, mark size=2.0pt, mark=*]
  table[row sep=crcr]{%
0	0.000	\\
1	0.054	\\
2	0.106	\\
3	0.156	\\
4	0.201	\\
5	0.241	\\
6	0.274	\\
7	0.300	\\
8	0.323	\\
10	0.411	\\
12	0.490	\\
14	0.591	\\
16	0.727	\\
18	0.847	\\
20	0.895	\\
22	0.911	\\
24	0.990	\\
25	1.082	\\
26	1.124	\\
27	1.049	\\
28	1.024	\\
29	0.949	\\
};

\addplot [color=blue, line width=0.1pt, mark size=2.0pt, mark=triangle]
  table[row sep=crcr]{%
0.954	0.052	\\
1.971	0.108	\\
2.926	0.156	\\
3.944	0.205	\\
4.962	0.237	\\
5.980	0.273	\\
6.936	0.297	\\
7.954	0.337	\\
8.972	0.373	\\
9.926	0.421	\\
10.945	0.457	\\
11.963	0.493	\\
12.918	0.533	\\
13.934	0.618	\\
14.951	0.682	\\
15.968	0.746	\\
16.921	0.834	\\
17.875	0.891	\\
18.893	0.931	\\
19.912	0.955	\\
20.930	0.983	\\
21.950	0.995	\\
22.905	1.027	\\
23.924	1.035	\\
24.877	1.119	\\
25.893	1.195	\\
26.910	1.275	\\
27.935	1.147	\\
28.957	1.110	\\
};

\addplot [color=red, line width=0.1pt, mark size=2.0pt, mark=square]
  table[row sep=crcr]{%
0.000	0.000	\\
0.978	0.057	\\
1.957	0.109	\\
3.000	0.161	\\
3.978	0.209	\\
5.022	0.248	\\
6.000	0.279	\\
6.978	0.305	\\
7.957	0.327	\\
9.000	0.374	\\
10.044	0.418	\\
11.022	0.457	\\
12.000	0.501	\\
12.978	0.553	\\
14.022	0.614	\\
15.000	0.688	\\
15.978	0.762	\\
17.022	0.818	\\
18.000	0.857	\\
19.044	0.883	\\
20.022	0.888	\\
21.000	0.888	\\
22.044	0.888	\\
22.957	0.949	\\
24.065	1.058	\\
24.978	1.136	\\
26.022	1.153	\\
27.065	1.049	\\
28.044	0.993	\\
29.022	1.093	\\
};

\addplot [color=violet, line width=0.1pt, mark size=2.0pt, mark=star]
  table[row sep=crcr]{%
1.984	0.103	\\
3.968	0.196	\\
5.952	0.265	\\
7.996	0.307	\\
9.980	0.397	\\
12.024	0.480	\\
12.986	0.494	\\
14.008	0.603	\\
15.030	0.573	\\
15.992	0.757	\\
17.014	0.810	\\
18.036	0.844	\\
18.998	0.861	\\
20.020	0.861	\\
21.042	0.810	\\
22.004	0.894	\\
23.026	0.961	\\
24.048	1.045	\\
25.010	1.109	\\
26.032	1.120	\\
27.054	0.975	\\
28.016	1.131	\\
29.038	1.070	\\
};

\legend{present study, Liu et al. \cite{Liu}, Kurtulus \cite{Kurtulus}, Khalid and Akhtar \cite{Khalid}};

\end{axis}
\end{tikzpicture}%
	\caption{Lift coefficient as a function of the angle of attack, at Re = 1000. Present results are compared with those from the numerical works of Liu et al. \cite{Liu} (triangles), Kurtulus \cite{Kurtulus} (squares) and Khalid and Akhtar. \cite{Khalid} (stars). \label{fig:NACA_CL}}
\end{figure}
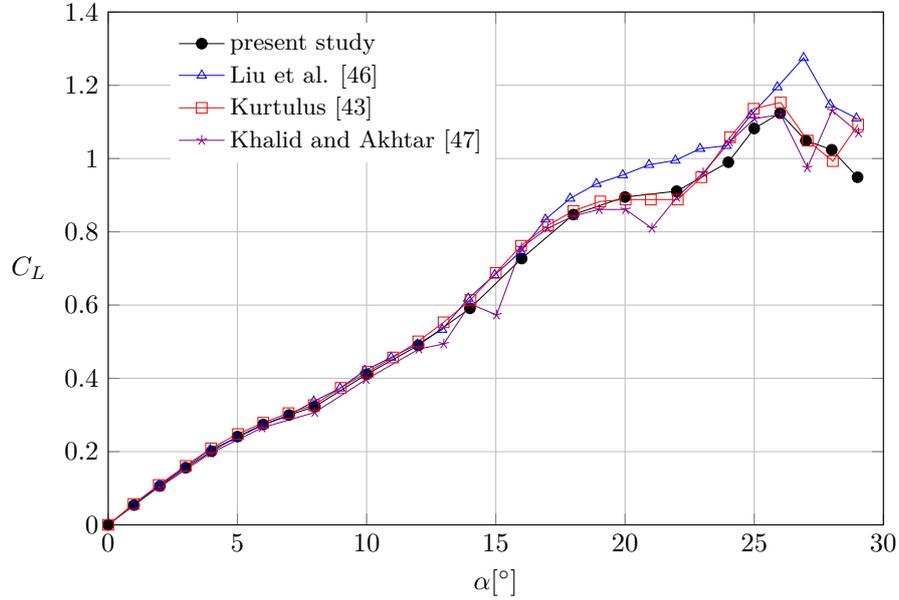

\begin{figure}[H]
	\centering 
	\setlength\figureheight{0.4\columnwidth}
	\setlength\figurewidth{0.6\columnwidth}
	\input{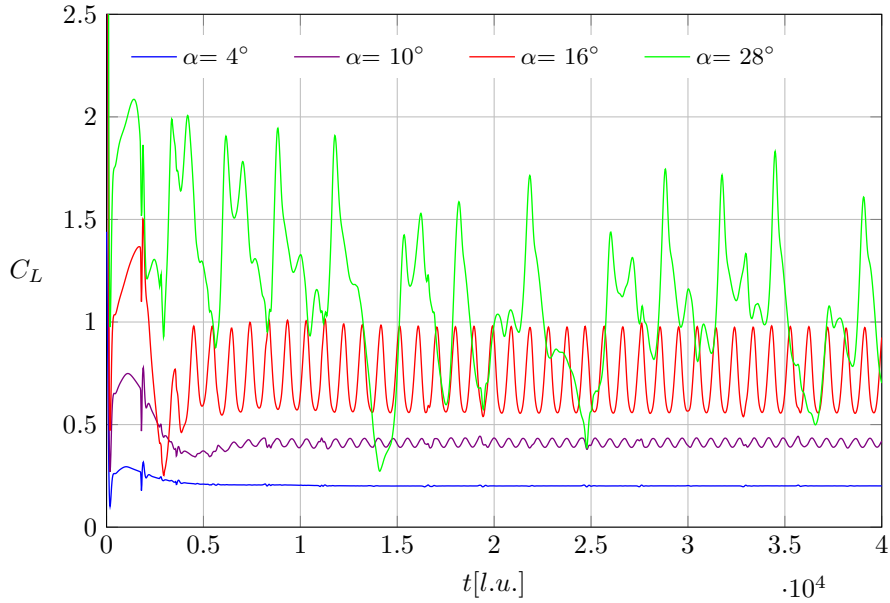}
	\caption{Time-history of lift coefficient for $\alpha=4^{\circ}$, $10^{\circ}$, $16^{\circ}$ and $28^{\circ}$. \label{fig:timeCL_NACA}}
\end{figure}

Overall, a good agreement between the solution obtained by HLBM and values from literature is achieved. In particular, the lift coefficient calculated by Liu et al. seems to be slightly overestimated at higher angles of attacks while the values computed by Kurtulus and by Khalid and Akhtar oscillates after a critical values is reached.
\\
The phenomenon of stall associated with airfoils is caused by a massive flow separation, which leads to a sudden drop in lift when the angle of attack is slightly increased beyond a critical value. This behavior seems to be properly captured by our numerical simulations. In particular, in this analysis, the numerical results suggest that stall occurs for a critical value of $\alpha$ around $26^{\circ}$, corresponding to a sharp drop in lift forces acting on the airfoil.
\\
Moreover, Figure \ref{fig:NACA_CD} shows that the drag coefficient remains fairly constant until a particular value of angle of attack is reached. This value corresponds, approximatively, to $\alpha=8^{\circ}$, when vortex shedding starts to appear. For angles of attack greater than $\alpha=8^{\circ}$, the drag force increases with $\alpha$, as expected.
\\
For the drag coefficient, we observe a good agreement between our findings and those available in the literature in the entire range of angles of attack analyzed. However, we notice a slight discrepancy at $\alpha=29^{\circ}$. Our analysis suggests that, for this particular case, the drag coefficient is close to that at $26^{\circ}$, corresponding to the critical value, while Kurtulus obtains a significant higher value. This difference is however acceptable, and can be attributed to the irregular behavior of the flow in this regime.
\\
Figure \ref{fig:timeCL_NACA} depicts the time history of the lift coefficient.
\\

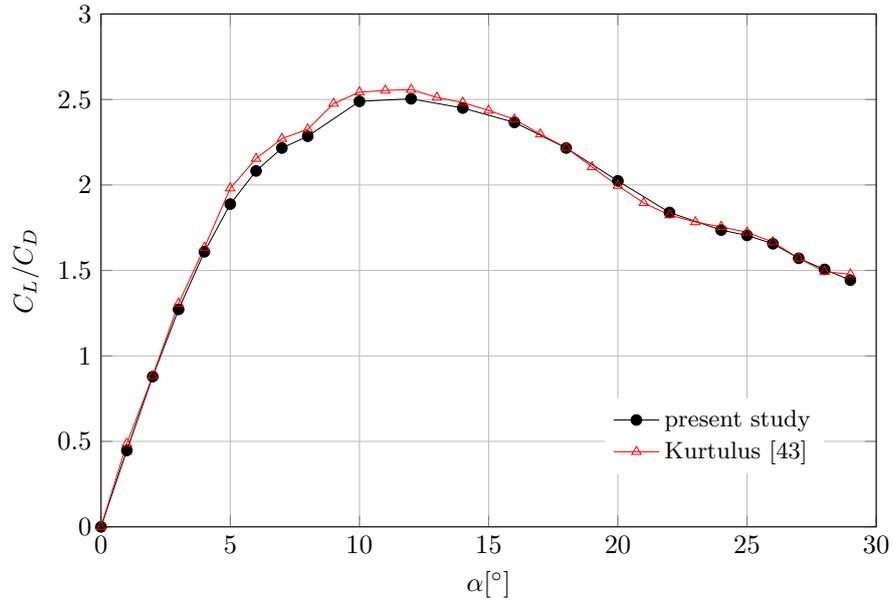
\begin{figure}[H]
	\centering 
	\setlength\figureheight{0.4\columnwidth}
	\setlength\figurewidth{0.6\columnwidth}
	\begin{tikzpicture}

\begin{axis}[%
width=\figurewidth,
height=\figureheight,
scale only axis,
xmin=0,
xmax=30,
xtick={0, 5, 10, 15, 20, 25, 30},
xlabel={$\alpha [^{\circ}]$},
xmajorgrids,
ymin=0,
ymax=3,
ytick={0, 0.5, 1, 1.5, 2, 2.5, 3},
ylabel={$C_L$/$C_D$},
ymajorgrids,
legend style={draw=none, at={(0.65,0.25)},anchor=north west,legend cell align=left,font = \small}
]

\addplot [color=black, line width=0.1pt, mark size=2.0pt, mark=*]
  table[row sep=crcr]{%
0	0.000	\\
1	0.447	\\
2	0.879	\\
3	1.272	\\
4	1.609	\\
5	1.888	\\
6	2.082	\\
7	2.216	\\
8	2.285	\\
10	2.489	\\
12	2.504	\\
14	2.450	\\
16	2.366	\\
18	2.216	\\
20	2.024	\\
22	1.839	\\
24	1.736	\\
25	1.705	\\
26	1.656	\\
27	1.571	\\
28	1.506	\\
29	1.443	\\
};

\addplot [color=red, line width=0.1pt, mark size=2.0pt, mark=triangle]
  table[row sep=crcr]{%
0	0.000	\\
1	0.487	\\
2	0.883	\\
3	1.307	\\
4	1.634	\\
5	1.981	\\
6	2.154	\\
7	2.271	\\
8	2.326	\\
9	2.476	\\
10	2.543	\\
11	2.553	\\
12	2.558	\\
13	2.512	\\
14	2.482	\\
15	2.436	\\
16	2.384	\\
17	2.297	\\
18	2.218	\\
19	2.105	\\
20	1.996	\\
21	1.895	\\
22	1.824	\\
23	1.783	\\
24	1.755	\\
25	1.723	\\
26	1.666	\\
27	1.571	\\
28	1.493	\\
29	1.478	\\
};

\legend{present study, Kurtulus \cite{Kurtulus}};

\end{axis}
\end{tikzpicture}%
	\caption{Lift-to-drag ratio vs angle of attack $\alpha$: comparison between present results and those reported in \cite{Kurtulus}.  \label{fig:CL_vs_CD}}
\end{figure}

The numerical results demonstrate that, below an angle of attack of 8 degrees, the lift coefficient converges to a constant value. This behavior corresponds to a steady state solution, since no fluctuations in the wake of the airfoil occur.
\\
However, due to the occurring of a trailing-edge vortex shedding, at $\alpha=8^{\circ}$ the lift coefficient begins to oscillate periodically. Above this particular value of $\alpha$ the flow becomes unsteady and the amplitude of the lift coefficient differs from zero. As a matter of example, in Figure \ref{fig:timeCL_NACA}, for $\alpha$ equal to 10 and 16 degrees, the lift coefficient is shown to exhibit a periodic trend.
\\
Moreover, it is possible to note that, as the angle of attack is increased, the amplitude of oscillation of the lift coefficient becomes larger. 
\\
The angle of attack for which the stall phenomenon occurs is known as stall angle. At this critical value, the developing of leading-edge vortices on the surface of airfoil and the massive flow separation lead to irregular oscillations of lift coefficient.
\\
Finally, the aerodynamic characteristic $C_L/C_D$ against the angle of attack is plotted in Figure \ref{fig:CL_vs_CD}.

From the analysis, it emerges that the maximum aerodynamic efficiency occurs between 10 and 12 degrees, since the ratio $C_L/C_D$ reaches a pick. Present computational results perfectly fit numerical literature data \cite{Kurtulus}.

\subsection{Effect of the Reynolds number}
\begin{figure}[H]
\centering
{\includegraphics[width = 0.4\columnwidth]{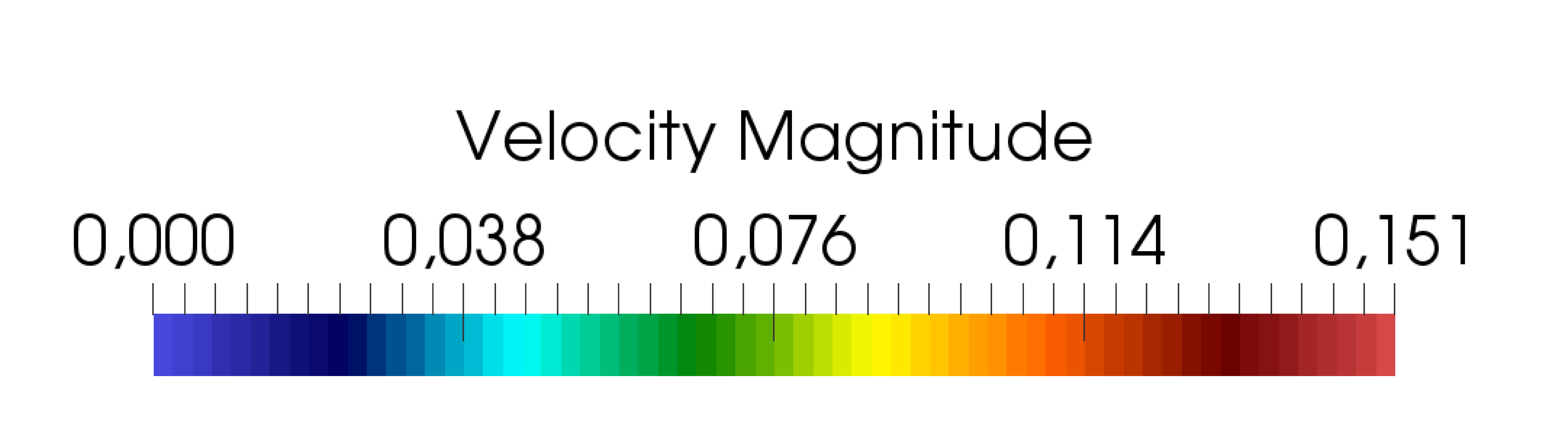}}
{\includegraphics[width = 0.6\columnwidth]{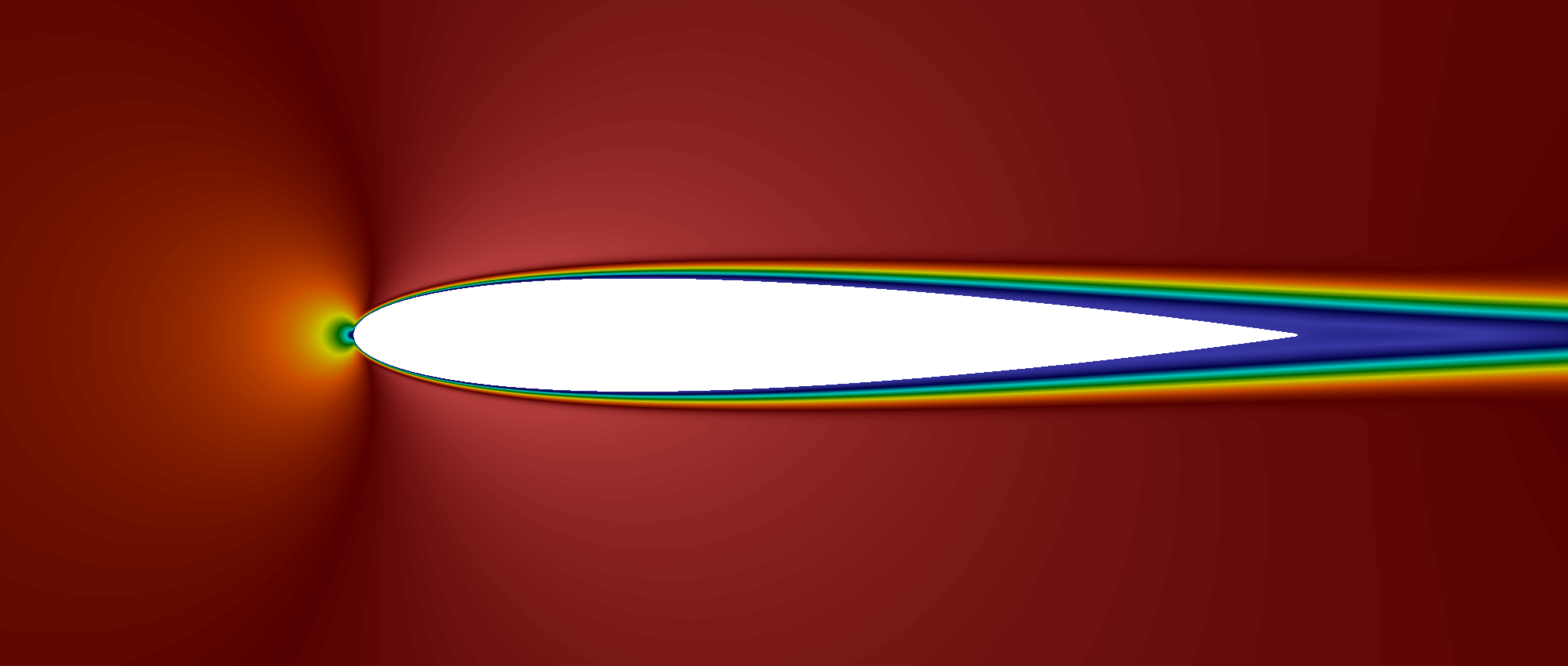}}
\caption{Instantaneous velocity field at Re = 10000.
\label{fig:velField_1000}}
\end{figure}

\begin{figure}[H]
\centering
{\includegraphics[width = 0.4\columnwidth]{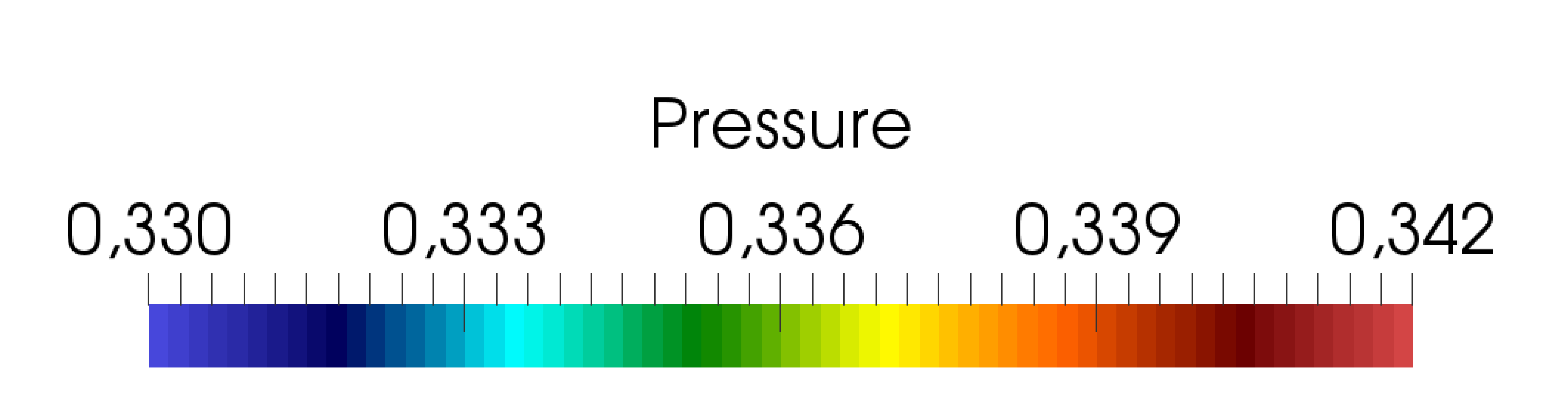}}
{\includegraphics[width = 0.6\columnwidth]{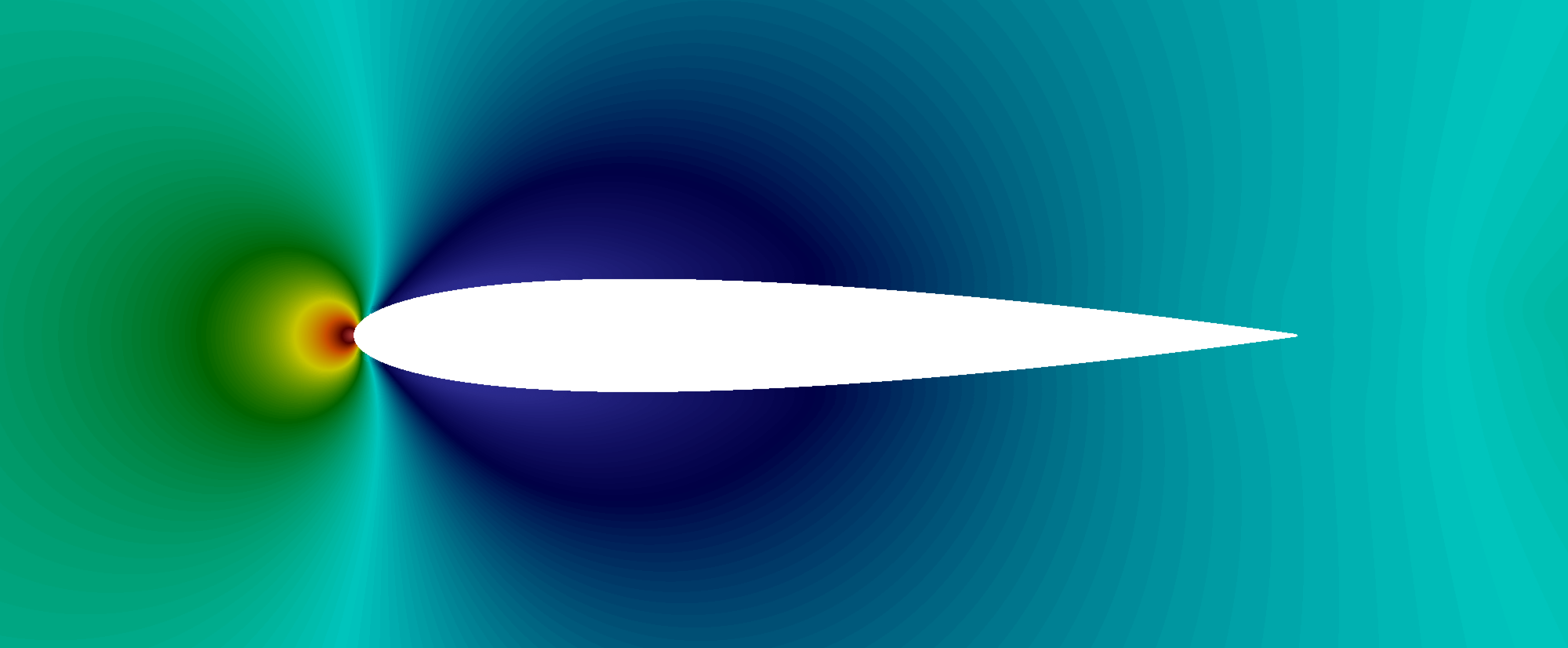}}
\caption{Instantaneous pressure field at Re = 10000.
\label{fig:preField_1000}}
\end{figure}

In this section we discuss the flow fields computed for Re = 2000, 5000 and 10000 at an angle of attack of $\alpha=0^{\circ}$.
\\
The same computational domain configuration illustrated in Figure \ref{fig:domain_NACA} is employed, while the chord length is taken to be equal to 1024 lattice units, referring to the finest structured refinement level. In this case, the hybrid mesh is composed of approximately $12.6\cdot10^6$ structured nodes and $8.8\cdot10^4$ unstructured nodes.
\\
Figures \ref{fig:velField_1000} and \ref{fig:preField_1000} show velocity and pressure fields for Re = 10000.


In order to validate the solution, a comparison between the results obtained by HLBM simulations and those calculated from \textit{XFOIL} is made. 
Table \ref{tab:dragcoeff_comparison} reports the values for the drag coefficient as a function of Reynolds number.

\begin{table}[h!]
\centering
\begin{tabular}{ c  c  c  c  c  c }
\hline
\multirow{2}{*}{} &\multirow{2}{*}{Re} & \multirow{2}{*}{} & \multicolumn{2}{c}{$C_D$} & \multirow{2}{*}{} \\
&       & & HLBM  & XFOIL & \\ \hline
&  1000 & & 0.119 & 0.119 & \\
&  2000 & & 0.084 & 0.084 & \\
&  5000 & & 0.052 & 0.054 & \\
& 10000 & & 0.037 & 0.040 & \\
\hline
\end{tabular}
\vspace{5mm}
\caption{Drag coefficients obtained by HLBM in comparison with values calculated by XFOIL for NACA0012 airfoil at $\alpha=0$.
\label{tab:dragcoeff_comparison}}
\end{table}
We note that the two solutions agree with each other very well for the entire range of Reynolds number investigated.

\subsection{Numerical performances}
In order to estimate the gain in efficiency of the HLBM with respect to the standard LBM, we consider an hybrid mesh which is composed of a uniformly spaced region and the same unstructured mesh employed in the analysis described in \ref{AoAstudy}. Also, we consider the same domain size and set of numerical parameters.
A practical way to quantify the difference of computational cost between HLBM and LBM is provided by the following definition of speedup (gain) indicator, introduced in \cite{DiIlio2017}:
\begin{equation}
G = \frac{T_\textbf{h}}{T_\textbf{eq}}=\frac{1+\chi s'n}{g}.
\label{eq:speedup}
\end{equation}
In equation \ref{eq:speedup}, $T_h$ is the wall-clock time per algorithm iteration associated to the HLBM, while $T_{eq}$ represents the wall-clock time per algorithm iteration with reference to an equivalent LBM implementation on uniform spaced mesh, that is a standard LBM with similar resolution around the solid body. The parameter $\chi$ in equation \ref{eq:speedup} indicates the ratio of the number of unstructured nodes to that of structured ones, $g$ is the ratio of the number of structured nodes which would be required by an equivalent LBM to the number of structured nodes actually used in the HLBM, while $s'$ represents the ratio of the processing speed for the standard LBM to the specific processing speed of the unstructured method. The unstructured finite-volume scheme is computationally more demanding than the standard LBM one, with typical values of $s'$ ranging between 3 and 5, as confirmed also in the present study.
By considering that the resolution of the unstructured mesh in proximity of the solid body is around 0.8 lattice units, we find $g\simeq1.6$. Also, for the present hybrid mesh, we find $\chi\simeq1.5\cdot10^{-4}$. Moreover, in all the performed simulations, we set the number of sub-iterations required by the unstructured method as $n=10$. By setting such a set of constants in equation \ref{eq:speedup}, we obtain $G\simeq0.63$. This value indicates that, for the case under consideration, the hybrid method runs faster than an equivalent standard LBM since, by definition, a value of $G$ less than 1 denotes the regime where the computational cost of HLBM is lower than that of standard LBM, when a similar resolution around the body is considered.
\\
To conclude, the numerical efficiency of the HLBM is, in general, higher than the one corresponding to the standard LBM. This is due to the presence of the unstructured mesh that, despite being computationally more expensive to solve than the uniform spaced one which is inherent to standard LBM, represents a small portion of the overall discretized domain. 
Therefore, the unstructured method provides a very flexible strategy for local grid refinement with major potential for multiscale applications.

\section{Conclusions}

The HLBM capability to properly predict the fluid flow behavior of a complex case study, such as that over an airfoil in static stall, has been assessed through this work.
The analysis, performed for a NACA 0012 airfoil at relatively low Reynolds numbers and different angles of attack, shows that the hybrid method is able to provide accurate results.
\\
Boundary layer separation, static stall, as well as the other physical phenomena involved, were captured by the numerical simulations. In particular, we found that, for a Reynolds number equal to $10^3$, the vortex shedding regime starts at an angle of attack of $8^{\circ}$, while the critical angle of attack, for which the stall phenomenon occurs, is around $26^{\circ}$. For angles of attack greater than this value, we observe a sharp drop in lift forces acting on the airfoil. Such results are in line with those available in literature. Overall, the involved fluid flow phenomena seems to be well predicted and the computation of drag and lift forces acting on the body, in the whole range of parameters analyzed, leads to satisfactory results.
\\
This achievement strengthens the robustness of HLBM thus projecting it as a potential tool for dealing with complex geometries. In fact, the presence of a body-conformed unstructured mesh allows high accuracy solution of the boundary layer regions, since a precise definition of the solid body profile as well as a flexible refining of the discretized domain is achievable. Moreover, thanks to the feature of LBM of allowing a local computation of the stress tensor at nodes of the computational domain, the HLBM prevents from the need of implementing boundary-fitting procedures for the computation of hydrodynamic forces on solid bodies. This property, which is partially lost in traditional lattice Boltzmann approaches, is here fully retained. Further, the method presents high computational efficiency, since the unstructured model is confined to a relatively small portion of the computational domain.
\\
These features, exploited throughout this study, are instrumental to the implementation of schemes for solving multi-scale problems.
The HLBM can be regarded also as an interesting alternative to traditional computational fluid dynamic approaches as well as to other off-lattice Boltzmann methods to the simulation of turbulent flows.
In fact, to properly capture all the main features of a fully developed turbulent flow, high resolution is required in those region of the fluid domain which are characterized by the smallest scales.
On the other hand, the numerical efficiency of the method plays a key role, therefore the geometric flexibility of a non-uniform mesh is often desirable.
To extend the applicability of the HLBM to turbulent flows, the implementation of a three-dimensional scheme is required. Extensions of the unstructured method used in this work are already present in literature \cite{Rossi}. Therefore, further studies will be aimed to develop suitable interpolation schemes for three-dimensioanl interfaces.

\section*{Acknowledgments}
The numerical simulations were performed on \textit{Zeus} HPC facility, at the University of Naples "Parthenope"; \textit{Zeus} HPC has been realized through the Italian Government Grant PAC01$\_$00119  \textit{MITO - Informazioni Multimediali per Oggetti Territoriali}, with Prof. Elio Jannelli as the Scientific Responsible.
\\
One of the authors (Sauro Succi) was partially supported by the European Research Council under the European Union's Horizon 2020 Framework Programme (No. FP 2014-2020) ERC Grant Agreement No. 739964 (COPMAT).

\pagebreak


\end{document}